\documentclass[final,3p,times]{elsarticle}
\usepackage{graphicx}
\usepackage{amssymb}
\usepackage{amsmath}
\usepackage{hyperref}
\usepackage{color}
\usepackage{dsserif}
\usepackage[normalem]{ulem}
\journal{Physics Letters A}
\DeclareUnicodeCharacter{2212}{-}

\usepackage{hyperref}
\usepackage[T1]{fontenc} 
\journal{Computer Physics Communications}
\begin{document}

\begin{frontmatter}

\title{OpenMP Fortran programs for  rotating dipolar Bose-Einstein condensates }

\author[scl]{Denis Mujo}
\ead{denis.mujo@ipb.ac.rs}

\author[scl]{Du\v{s}an Vudragovi\'c}
\ead{dusan.vudragovic@ipb.ac.rs}

\author[bdu,bdu2]{Paulsamy Muruganandam}
\ead{anand@bdu.ac.in}

\author[ift]{Sadhan K. Adhikari\corref{author}}
\ead{sk.adhikari@unesp.br}
\cortext[author]{Corresponding author.}

 \address[scl]{Scientific Computing Laboratory, Center for the Study of Complex Systems, Institute of Physics Belgrade, University of Belgrade, Pregrevica 118, Belgrade, 11080, Serbia}

\address[bdu]{Department of Physics,  Bharathidasan University, Palkalaiperur Campus, Tiruchirappalli 620024, Tamilnadu, India}

\address[bdu2]{Department of Medical Physics,  Bharathidasan University, Palkalaiperur Campus, Tiruchirappalli 620024, Tamilnadu, India}

\address[ift]{Instituto de F\'{\i}sica Te\'{o}rica, UNESP -- Universidade Estadual Paulista, 01.140-70 S\~{a}o Paulo, S\~{a}o Paulo, Brazil}

\begin{abstract}

In this paper we present Open Multi-Processing (OpenMP) Fortran 90/95 programs to solve the  Gross-Pitaevskii equation  for a rotating dipolar Bose-Einstein condensate (BEC)  in two and three dimensions, which is a new version of our previous published programs for a dipolar Bose-Einstein condensate without rotation.  After the recent experimental study of a rotating  dipolar BEC [L. Klaus et al., Nature Phys. 18, 1453 (2022)], the present programs will be useful tools  for   related theoretical investigation.
The algorithm used is the split-step semi-implicit Crank-Nicolson scheme for imaginary- and real-time propagation to obtain stationary states and BEC dynamics, respectively,  as in the previous version [L. E. Young-S. et al., Comput. Phys. Commun. 286 (2023) 108669].  
\end{abstract}


\end{frontmatter}

\begin{small}
\noindent
{\bf New version program summary}

\noindent\vspace*{-2mm}\\
{\em Program title:} DBEC-ROT-PROGRAM, a program package containing programs imre3d-th.f90, 
imre2dXY-th.f90,     with fftw3.f03 and fftw3.mod.

\noindent\vspace*{-2mm}\\
{\em CPC Library link to program files:} \url{} 

\noindent\vspace*{-2mm}\\
{\em Program obtainable from:} CPC Program Library, Queens University, Belfast, N. Ireland.

\noindent\vspace*{-2mm}\\
{\em Licensing provisions:} Apache License 2.0

\noindent\vspace*{-2mm}\\
{\em Programming language:} Open Multi-Processing (OpenMP)  Fortran 90/95. The program is tested with the GNU, Intel,  and Oracle (former Sun) compilers.

%\noindent\vspace*{-2mm}\\
%{\em Supplementary material}: File Supp.pdf gives additional details about the new program version and the %underlying physical system.

\noindent\vspace*{-2mm}\\
{\em Journal Reference of previous version}: {Comput. Phys. Commun. 286   (2023) 108669.} {https://doi.org/10.1016/j.cpc.2023.108669}

\noindent\vspace*{-2mm}\\
{\em Does the new version supersede the previous version?}:  Yes 
%{\color{green}Only partially. The program spin-SO-rot-imre2d-omp.f90 supersedes spin-SO-imre2d-omp.f90, while the one-dimensional program is not part of this package.}
 
% \noindent
\vspace*{+2mm}
 {\it Nature of problem:}
The present  OpenMP Fortran 90/95 programs solve the time-dependent nonlinear partial differential Gross-Pitaevskii (GP) equation for a trapped rotating dipolar Bose-Einstein condensate (BEC) in  two (2D), and three  (3D) spatial dimensions.  In  3D the provision for including the higher-order repulsive  Lee-Huang-Yang interaction   is accommodated.

 \vspace*{+2mm}
{\em Solution method:}
We employ the split-step Crank-Nicolson scheme to discretize the time-dependent GP equation in space and time. The discretized equation is then solved by imaginary- or real-time propagation,   employing adequately small space and time steps, to yield the solution of stationary and non-stationary problems, respectively.

 \vspace*{+2mm}
{\em Reason for new version:}
 Previously  published Fortran  \cite{bec2009}  and C \cite{bec2012}programs and their OpenMP  and Cuda versions \cite{xxx,bec2016ompmpinondipC,dbec2016alldipC,dbec2016cudadipC} for solving the  time-dependent GP
equation  
have become useful tools in the study of statics and dynamics of a BEC. These programs have been  extended to the more complex scenario
of dipolar atoms \cite{dbec2015},   and rotating consensates \cite{vor-lat}. 
The available  Fortran 90/95 and C programs \cite{bec2009,bec2012,dbec2015} for the solution of the GP equation for a dipolar BEC  
  are enjoying widespread use in phenomenological application.  The same 
for the solution of the GP equation of  a rotating BEC have played  a fundamental role in the  theoretical study of the formation of vortices in a rotating BEC.   Although,  the  experimetal realization of a  rotating nondipolar  BEC  in a controlled fashion dates back to 1999 \cite{dalibard}, the same of a rotating dipolar BEC has  been possible only recently in  2022 \cite{ferlaino}.  In view of this experimental breakthrough,  the present  OpenMP  Fortran 90/95 programs for the solution of the time-dependent GP equation for a rotating dipolar BEC are of paramount interest. 
 
 \vspace*{+2mm}
{\em Summary of revisions:} 
The program package DBEC-ROT-PROGRAM contains the 3D program  imre3d-th.f90 and  2D program  
imre2dXY-th.f90 in    the directory src, as well as the files makefile, makefile-mur,  readme.txt, and
   readme-fftw.txt. The directory output contains examples of matching outputs of imaginary- and real-time propagation programs in sub-directories with a generic name, e.g.  imag3d, real2dXY, etc. These results are not fully  converged numerically. The reader is requested to read the file  readme.txt, which gives instruction to compile and run the programs   by the command make and also to install the fast Fourier transformation (FFT) routine in Linux/Unix operating systems.
The reader is advised to consult Ref. \cite{dbec2015} for details.  The 3D and 2D programs are OpenMP versions leading to a significant reduction in execution time in a multicore processor.  
    The makefile allows an automated compilation of the programs using different supported compilers (gfortran GNU, ifx Intel, macOS) by a simple make command, as in Ref.~\cite{dbec2015}. (However, the 1D and 2D dipolar programs in previous publications \cite{dbec2016alldipC,dbec2016cudadipC,dbec2015} erroneously used the 3D nonlinearities, in place of corresponding 1D and 2D nonlinearities,  in numerical calculations. The corresponding arXiv archive versions do not have these errors.)     All input parameters are placed in the beginning of each program in MODULEs GPE$\_$DATA and COMM$\_$DATA. Most of these input parameters are the same as in Ref. \cite{luis}. The new parameter XDIPOLAR (=0) bipasses the dipolar part to give the result for a nondipolar condensate. As in Ref. \cite{vor-lat}, the parameter FUNCTION selects the type of initial Gaussian function (with or without a vortex), and  the parameter RANDOM  (=1)  includes an arbitrary phase  on the initial Gaussian function to facilitate the creation of vortices. The parameter LHY (=1) allows the possibility to include  the Lee-Huang-Yang interaction in 3D.  The imaginary-time programs implement the calculation starting from an initial analytic Gaussian wave function using a non-zero value of the parameter NSTP, whereas the real-time programs use the converged solution of the imaginary-time programs as the initial state of calculation employing NSTP = 0. The FFT algorithm works faster and more efficiently when we take the number of space grid points  along $x$, $y$ and $z$ directions $-$   NX, NY in 2D  and NX, NY, NZ in 2D and 3D $-$ in powers of 2, e.g. $2^n$ with $n$ an integer, and should be so chosen for reducing the execution time. The operational scheme for running the codes is identical  to that in Ref. \cite{dbec2015}.

 \end{small}
 
\section*{CRediT authorship contribution statement}
Denis Mujo: Writing – review and editing, Validation, Software, Investigation, Formal analysis; Dušan Vudragovic: Writing – original
draft, Visualization, Validation, Supervision, Software, Investigation,
Conceptualization; Paulsamy Muruganandam: Writing – review and 
editing, Validation, Supervision, Software, Investigation, Conceptualization; Sadhan K. Adhikari: Writing – review and  editing, Visualization,
Validation, Supervision, Software, Investigation, Conceptualization.

\section*{Data availability}
Data used in  this paper  are included in the program folder 
DBEC-ROT-PROGRAM.

\section*{Declaration of competing interest}
The authors declare that they have no known competing financial
interests or personal relationships that could have appeared to influence
the work reported in this paper.
\section*{Acknowledgments}
\noindent
D. M. and D. V. acknowledge support provided by the Institute of
Physics Belgrade, through the grant by the Ministry of Science, Technological Development, and Innovations of the Republic of Serbia. The
work of P. M. is supported by the Ministry of Education-Rashtriya
Uchchatar Shiksha Abhiyan (MoE RUSA 2.0): Bharathidasan University–
Physical Sciences. S. K. A. acknowledges support by the Conselho Nacional de Desenvolvimento Científico e Technológico (National Council for Scientific and Technological Development, Brazil) grant
303885/2024-6. Numerical simulations were run on the PARADOX supercomputing facility at the Scientific Computing Laboratory, National
Center of Excellence for the Study of Complex Systems, Institute of
Physics Belgrade.


\begin{thebibliography}{13}

\bibitem{bec2009}
P. Muruganandam, S.~K. Adhikari, Fortran programs for the time-dependent Gross-Pitaevskii equation in a fully anisotropic trap, Comput. Phys. Commun. 180 (2009) 1888; arXiv:0904.3131.



\bibitem{bec2012}
D. Vudragovi\'c, I. Vidanovi\'c, A. Bala\v{z}, P. Muruganandam, S. K. Adhikari, C programs for solving the time-dependent Gross-Pitaevskii equation in a fully anisotropic trap,
Comput. Phys. Commun. 183 (2012) 2021; arXiv:1206.1361.


 \bibitem{xxx}
L. E. Young-S., P. Muruganandam, S. K. Adhikari, V. Loncar, D. Vudragovi\'c, Antun Bala\v z, OpenMP GNU and Intel Fortran programs for solving the time-dependent Gross-Pitaevskii equation,
Comput. Phys. Commun. 220 (2017) 503; arXiv:1709.04423. 


\bibitem{bec2016ompmpinondipC}
B. Satari\'c, V. Slavni\'c, A. Beli\'c, A. Bala\v z, P. Muruganandam, and S. K. Adhikari, Hybrid OpenMP/MPI programs for solving the time-dependent Gross-Pitaevskii equation in a fully anisotropic trap,
Comput. Phys. Commun. 200 (2016) 411; arXiv:1601.04641.


 
\bibitem{dbec2016alldipC}
V. Lon\v{c}ar, L.E. Young-S., S. \v Skrbi\'c, P. Muruganandam, S.K. Adhikari, A. Bala\v{z}, OpenMP, OpenMP/MPI, and CUDA/MPI C programs for solving the time-dependent dipolar Gross-Pitaevskii equation, Comput. Phys. Commun. { 209} (2016) 190; arXiv:1610.05329. 


 
\bibitem{dbec2016cudadipC} 
V. Lon\v{c}ar, A. Bala\v{z}, A. Bogojevi\'{c}, S. \v{S}krbi\'{c}, P. Muruganandam, S.K. Adhikari, CUDA programs for solving the time-dependent dipolar Gross-Pitaevskii equation in an anisotropic trap, Comput. Phys. Commun. { 200} (2016) 406; arXiv:1601.04640. 

 


\bibitem{dbec2015}
R. Kishor Kumar, L.E. Young-S., D. Vudragovi\'{c}, A. Bala\v{z}, P. Muruganandam,  S.K. Adhikari, Fortran and C programs for the time-dependent dipolar Gross-Pitaevskii equation in an anisotropic trap, Comput. Phys. Commun. { 195} (2015) 117; arXiv:1506.03283.





\bibitem{vor-lat}
R. K. Kumar, V. Lon\v car, P. Muruganandam, S. K. Adhikari, A. Bala\v{z}, C and Fortran OpenMP programs for rotating Bose–Einstein condensates,  Comput. Phys. Commun. { 240} (2019) 74; arXiv:1906.06327. 
 

\bibitem{dalibard} K. W. Madison, F. Chevy, W. Wohlleben, J. Dalibard,
  Vortex formation in a stirred Bose-Einstein condensate, Phys.  Rev. Lett. 84 (1999)  806; 	arXiv:cond-mat/9912015.
 
 
 \bibitem{ferlaino}L. Klaus, T. Bland, E. Poli, C. Politi, G. Lamporesi, E. Casotti, R. N. Bisset, M. J. Mark, Francesca Ferlaino, Observation of vortices and vortex stripes in a dipolar condensate, Nature Phys. 18 (2022)  1453;  arXiv:2206.12265.

 \bibitem{luis}  L. E. Young-S., P. Muruganandam, A. Balaž, S. K. Adhikari, OpenMP Fortran programs for solving the time-dependent dipolar Gross-Pitaevskii equation, Comput. Phys. Commun. 286 (2023) 108669; arXiv:2301.09383.


\end{thebibliography}
\end{document}